# Learning Analytics Made in France: The METALproject

Armelle Brun, Geoffray Bonnin, Sylvain Castagnos, Azim Roussanaly, Anne Boyer

LORIA UMR 503 – Université de Lorraine

Campus Scientifique – 54506 Vandoeuvre les Nancy – Cedex

{armelle.brun,geoffray.bonnin, sylvain.castagnos, azim.roussanaly, anne.boyer}@loria.fr

## Abstract

This paper presents the METAL project, an ongoing French open Learning Analytics (LA) project for secondary school, that aims at improving the quality of the learning process. The originality of METAL is that it relies on research through exploratory activities and focuses on all the aspects of a Learning Analytics implementation.

This large-scale project includes many concerns, divided into 4 main actions. (1) data management: multi-source data identification, collection and storage, selection and promotion of standards, and design and development of an open-source Learning Record Store (LRS); (2) data visualization: learner and teacher dashboards, with a design that relies on the co-conception with final users, including trust and usability concerns; (3) data exploitation: study of the link between gaze and memory of learners, design of explainable multi-source data-mining algorithms, including ethics and privacy concerns. An additional key of originality lies in the global dissemination of LA at an institution level or at a broader level such as a territory, at the opposite on many projects that focus on a specific school or a school curriculum.

Each of these aspects is a hot topic in the literature. Taking into account all of them in a holistic view of education is an additional added value of the project.

## Keywords

Educational data, multi-source data, data collection and storage, teacher and learner dashboards, data mining.

## 1 Introduction

The domain of Learning Analytics (LA) has recently evolved from using data for predicting which learners are at risk of dropout/academic failure, to studying how to improve the learning experience and academic outcomes of all learners.

LA tools can help teachers improve their own practice. For example, LA can inform teachers about the quality of the pedagogical contents they use, the impact of the activities they propose and their assessment process (Jivet, Scheffel, Specht, & Drachsler, 2018). LA can also be used to help teachers monitor in real-time the performance of learners, or to adapt their teaching (Guo, J., Huang, X., & Wang, B., 2017).

LA tools can also help learners to get information about how they are progressing and what they need to do to meet their educational goals (Davis, S. K., Edwards, R. L., Miller, M., & Aragon, J., 2018). In other words, LA is an opportunity for learners to take control of their own learning, as it informs them about their involvement, activity, performance and progression.

Many works relate experiments about monitoring, predicting and visualizing learners' learning behaviour, in terms of motivation, work, engagement, etc. mainly at a subject level (Hu, X., Zhang, Y., Chu, S., &

Ke, X., 2016), but the issue of a global dissemination of LA at an institution level or at a broader level such as a territory or a study level is still a hot topic. It is the reason why the French Learning Analytics project called METAL, has been launched at the end of the year 2016, for a period of four years. The main objective of METAL is to embrace all the aspects of a LA institutional strategy, from data identification, collection and storage to the impact study. It also includes an algorithmic perspective: multi-source data mining, study the link between learners' gaze and memory, and a visualization perspective: teachers and learners enrolment and support through dashboards. METAL aims at proposing a generic framework to encourage the development and dissemination of LA tools in French secondary school (from K7 to K10), and a first experiment is currently made at the scale of the Lorraine regional territorial area.

In French secondary school, many actors are involved in the provision of the digital environment of learners: the regional education authority, the administrative region, the school, private editors, etc. and each of them collects, and owns, associated data (mainly logs, textual information, ratings, like/dislike, answers to quiz, assessments, etc.). In addition, the curriculum of each year is defined at a national level, but each school is free to select the textbooks and softwares that will be used. It results in a great diversity of educational contexts that have to be at least interoperable.

Managing these numerous educational contexts constitute a major challenge. There are many barriers in French secondary school that prevent data from being used effectively. Let us mention issues such as data availability, quality, ownership, scarcity, access, privacy, trackability or ethics. The first main question to answer is how to identify all the data available in any educational context, and how to support and secure their ethical use. To achieve this challenge, one concern of METAL is the selection and the promotion of standards.

METAL addresses these issues through a methodology to map the educational data in a specific context, the design of an open-source LRS (Learning Record Store), and the definition of a charter for a responsible and ethical use of educational data. A LRS is a central independent point to collect all event data from both the LMS (Learning Management System) and behavior sensors (Flanagan, B. & Ogata, H., 2018).

As feedback is a key element by which learners and teachers engage in a dialogue with the aim of improving the learning experience (Nicol, 2010), METAL also focuses on the design of dashboards with a co-design approach, in particular at the level of a course or of a set of courses. The first one provides teachers with pedagogical indicators on their learners and the second one provides learners with indicators on their learning behaviour. Teacher dashboards are designed to develop the capacity to visualize data just in time and with the adequate point of view for teachers to decide which actions are most appropriate (possibly helped by recommendations). The learner dashboard provides an insight into their behaviour, and recommendations adapted to their needs. Such a feedback is designed to facilitate improvements, reinforce academic performance, support engagement and increase motivation. A related question is the design of both dashboards. The well-known need to focus on actionable knowledge in LA also extends to the associated tools. Therefore, explicit actions will be recommended to support the learners.

Data mining techniques (Han, J., Pei, J., & Kamber, M., 2011) will be designed to showcase information, such as the relation between data (from the Virtual Learning Environment (VLE) for example, data available in the specific context of the secondary school, etc.) and educational outcomes. The resulting model is used in both dashboards. The mandatory need to form explainable models (for the final user) (Gunning, 2017) impacts the kind of data mining algorithms. The question addressed is how to design data mining algorithms, able to explain in an understandable fashion their results obtained on heterogeneous and multi-source data?

To summary, METAL is original in many ways: (1) METAL is a large-scale project that aims at a global dissemination of learning analytics at an institution level or at a broader level, (2) the focus of METAL are secondary school learners, (3) METAL regroups a wide range of stakeholders including researchers, teachers, learners, parents, institutions, etc., (4) METAL tackles with several issues related to data management, data visualization, data exploitation, (5) METAL focuses on both technical aspects and implementation with final users (mainly learners and teachers); co-design as well as ethics are at the core of the actions carried.
Although METAL is an ongoing project, first conclusions have been drawn and are presented in this paper, ongoing experiments are also introduced, as well as future works.

Management is an important dimension in this multifaceted project. Before starting the project, several meetings have been conducted to converge towards a common view of LA, the expected output, the impact on and the role of all the stakeholders, and the definition of a common roadmap. Throughout the entire project, all the stakeholders are involved, to be ensured that each of them fully agrees with the objective of each task and the way it is processed. This involvement is of primary importance to ensure the final adoption of all the outputs by each stakeholder.

This paper presents a global overview of METAL, and is organized as follows: Section 2 is dedicated to the problem of identifying, collecting and storing data. Section 3 presents the way teacher and learner dashboards are designed. Section 4 highlights two specific mining tasks. Section 5 concludes the presentation of this project.

## 2   Data Collection and Storage

As soon as METAL was drafted, it was observed that multiple data sources and data owners had to be considered to capture information: content editors, Internet providers, software owners, institutions, regional education authority, etc. Each data source shows its own complexity, at both technical and legal levels.
Considering all the sources, their heterogeneity and diversity, together with the need of interoperability, the approach of collecting data in a LRS aims to tackle a set of issues namely:
- considering the ethical and legal aspects of personal data collection in a global approach of ethics by design (Drachsler and Greller, 2016);
- structuring the data in a centralized way to avoid complex data collection, interpretation and formatting procedures for e-services;
- designing the outline of a generalizable standard at the level of French secondary school.

Building a multi source LRS can be considered as close to designing a data warehouse (Miloslavskaya and Tolstoy, 2016). In this sense, the approach adopted in METAL is inspired by the classic data mining methodology, based on the Cross Industry Standard Process for Data mining (CRISP-DM) (Chapman, Khabaza, Shearer, & CRISP-DM, 2000). Although this standard is quite old, it is still used as a reference by the community of data mining actors (Piatesky, 2014). Moreover, only the first three steps are concerned by the data collection and storage process, namely: (1) the business understanding, (2) the data understanding, (3) the data preparation. In the the specific context of METAL these steps could be reformulated respectively as (1) the stakeholder identification, (2) the data modeling, and (3) learning record store (LRS) designing.

In the first phase, the stakeholders are identified, as well as their respective roles and mutual interactions in terms of service providing and treatment responsibility. Figure 1 presents a schematic view of these stakeholders. It shows the main services likely to provide data for METAL:
- the Internet access, under the responsibility of the County Councils.
- the Virtual Desktop available for the learners, their parents, the teachers and the educational staff, funded by each County Council, whose deployment is under the responsibility of the Digital

Learning Service (DANE) and deployed by a private partner company that collects and stores data,
- the School Management Suite (including the School Life Application) used by the educational establishments; data is collected by either the Suite provider, or the secondary school;
- the Educational Resources Portal to access digital educational resources under the responsibility of the Ministry of Education (delegated to the Educational Resource Provider Service or BRNE). This portal is powered by different independent publishers of digital resources and textbooks, which collect data of learners' activity, and funded by the Ministry of Education for a national use.

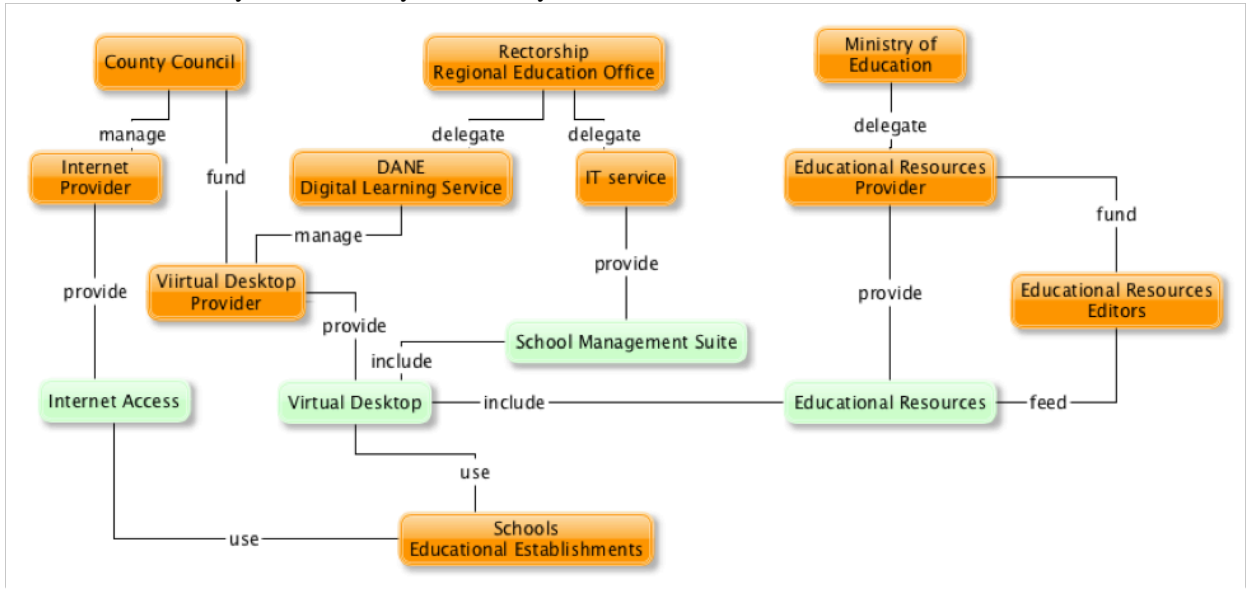

**Figure 1 Stakeholders in the METAL project.**

The second phase is related to data modelling. The main on-line services mentioned above contain or generate data that could be used in the different LA applications. In a bottom-up approach that starts from a mapping of available data, the goal is to combine and structure them in the formalism of a conceptual data model. An analysis of the data (and sources) available according to their nature is conducted:
- *Demographic* (about learners and their families): nationality, place and date of birth, address, genre, date of birth, the socio-professional category of the parents,
- *Curriculum*: the annual results and reports of training cycle, possibly associated with the educational establishments,
- *School life*: data available within the institution during the current school year, includes organization of the classes, learners' notebook, marks and reviews, their homework, their absences and incidents,
- *Educational resources*: the description (metadata) associated with pedagogical resources,
- *Learner activity*: data from the learners' online actions, whether they access an online service, or achieve an activity on the LMS or use a digital educational resource.

Once the data has been identified, they had to be structured in a conceptual formalism in a holistic user modeling perspective that puts the user entity at the center of the process (Musto, Semeraro, Lovascio, de Gemmis, & Lops, 2018). Figure 2 provides an overview of the resulting conceptual model obtained by the entity relationship (ER) modeling technique.

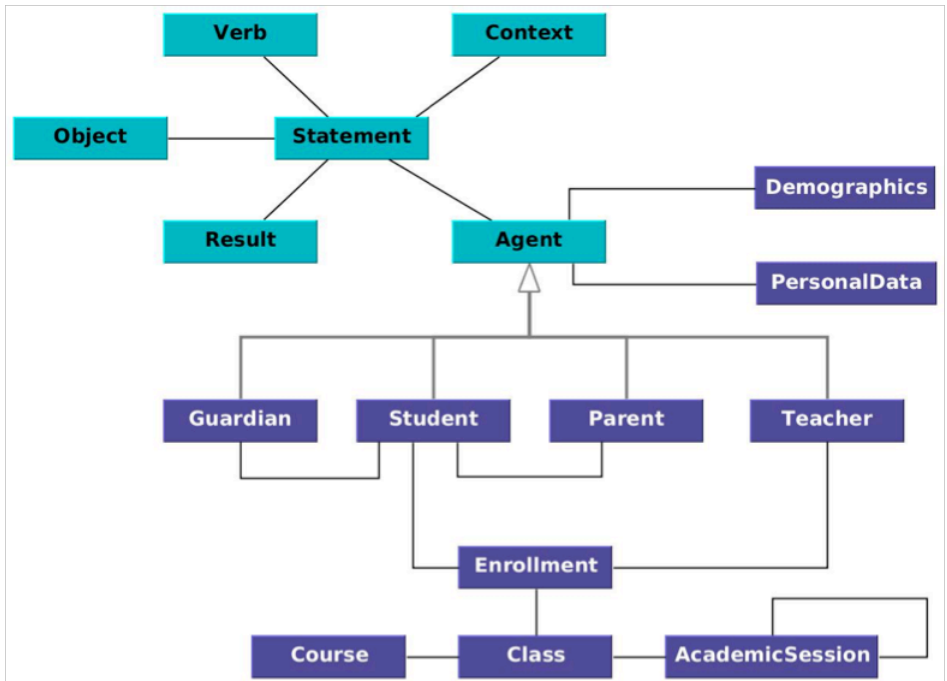
**Figure 2 Conceptual data model**

The third phase aims at facilitating both data warehouse supply and the data access by LA applications. The LRS therefore refers to a LA-specific data warehouse. Unlike the previous phase, a top-down approach is adopted, that consists in examining among the existing standards those that best match the data implementation plan chosen.
Digital learning offers many standards answering different issues. Two of them have been selected because they respond to the specific goal of METAL in a complementary way.

**eXperience API (xAPI)**
xAPI (Adlnet, 2017), designed by the Advanced Distributed Learning Initiative (ADL), presents, in the context of METAL, the following significant features: xAPI reports online learning activities, xAPI standardizes a set of endpoints, xAPI is adaptable to many environments, xAPI considers partially legal issues, xAPI is becoming a standard increasingly adopted by the international community of digital resource publishers.
However, xAPI only partly resolves the specific need for data storage. When referring to the conceptual diagram of Figure 2, xAPI covers only the green part. To consider the remaining part, the xAPI interface has to be extended to the LRS.

**OneRoster**
OneRoster (IMS Global, 2018) is a standard published by the IMS Global Learning Consortium. Implicitly, OneRoster relies on a data model that integrates both school organization, school life and educational activities, which makes it particularly interesting. However, the embedded model is particularly oriented towards a context of American institutions, whether in terms of organization or legal considerations. Consequently, it requires an adaptation to the French national system and the constraints of METAL, especially considering privacy on the one hand, and the coverage of complementary data (purple part in Figure 2) on the other hand.
Furthermore, if OneRoster offers the possibility of an exchange format between two information systems based on the creation of CSV standard files, it also defines an interface that uses fully described web services endpoints in the specifications of the standard. This last feature is particularly interesting because it technically allows a smart extension of xAPI, which is based on the same principle.

The LRS of the project is therefore the data warehouse designed in METAL, resulting from the joint implementation of both xAPI and OneRoster standards. Its architecture is shown in Figure 3.

The ongoing work aims to implement appropriate connectors for each identified data source in order to evaluate the performance of the LRS in real-life situations, based on the data collection from five schools selected for the experiment.

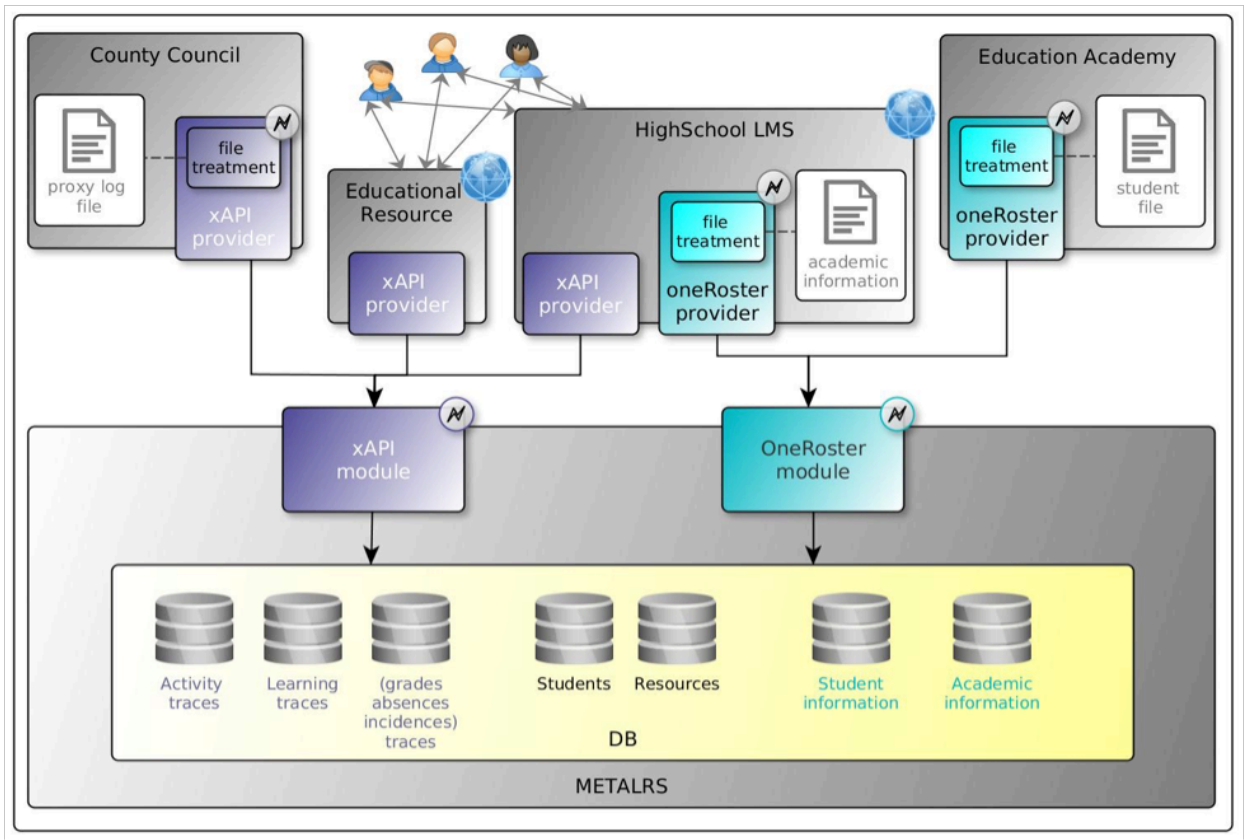

Figure 3 Architecture of the LRS of the project

## 3 Dashboards

METAL aims at providing teachers and learners with tools that support them in their respective tasks, including dashboards, which are the focus of this section. Such tools mainly rely on data gathered about learners, that result from the tracking of their online activities, and data collected from educational information systems (see section 2). This data can then be analyzed and mined with the goal to infer information about learners.

Dashboards are one way to visualize such information and are aimed to support teaching or learning activities. As defined in (Schwendimann, B. A. et al., 2017), Learning Analytics dashboards are "single displays that aggregate different indicators about learner(s), learning process(es) and/or learning context(s) into one or multiple visualisations".

The quality of a dashboard is mainly evaluated in terms of reliability of the information it contains and in terms of acceptance. Regarding reliability, the information displayed must not only be valid, but also up to date, coherent through time, etc. Regarding acceptance, the information displayed has to be understandable and understood by the final user, whoever he/she is, and whatever is his/her context, be compliant with privacy issues, etc. (Jivet et al., 2018).

In the literature, dashboards can be dedicated to either teachers (Kuzilek, Hlosta, Herrmannova, Zdrahal, & Wolff, 2015) or learners (Davis, S. K. et al., 2018), and in some cases to both of them (Millecamp, M., Gutierrez, F., Charleer, S., Verbert, K., & De Laet, T., 2018). Let us notice that most of the dashboards have been developed to support teachers (Fu, X., Shimada, A., Ogata, H., Taniguchi,Y., & Suehiro, D., 2017; Guo, J. et al., 2017), few of them were specifically developed to support learners. Thus, there is little known about the impact of learner-facing dashboards. One exception is the STELA project, in which student-facing dashboards are being developed to support students in their transition from secondary to higher education (Leitner, P. & Ebner, M., 2017). Differently to the STELA project, we put a strong emphasis on involving the stakeholders in the design of the dashboards, through co-design sessions.

Let us mention that LA raises pedagogical issues, that are discussed in many works. They can be related to teachers, such as the impact on their pedagogy (Heilala, 2018), how they incorporate LA into their teaching protocols (Kitto, Lupton, Davis, & Waters, 2016), etc. They can also be related to learners, where the impact on their learning is studied: performance (Mirriahi & Dawson, 2013), concentration (Lei, C.U., Hou, X., & Gonda, D., 2018), etc.

### *3.1 Dashboards for teachers*

By displaying (inferred) information about learners, dashboards can help the teacher in his new role as a coach. For example, (Xhakaj, Aleven, & McLaren, 2017) analyse the effect of a teacher dashboard on teachers and learners. They found that teachers highly adapted their lesson plan by exploiting 45% of the information provided by the dashboard.

The specific focus in METAL is to provide two types of views: a global view of the class and an individual view of each learner. Both views are designed to focus on the relationship between the effort of the learners and their engagement. Moreover, the individual view provides engaging recommendations of activities that are to be validated by the teacher before being forwarded to the learner. One key challenge of dashboards is their design and evaluation (Jivet et al., 2018). Therefore, in order to create such a dashboard, a series of co-design sessions were performed with a group of 13 teachers, in collaboration with the e-FRAN E-TAC project. Two sessions were set up: (1) a focus group to gather information related to the actual needs and difficulties of the teachers and (2) a session of layout prototyping using paper material.

The focus group session allowed us to make the following observations. First, teachers all agreed that such a dashboard could help them saving some time in analyzing learner outcomes. The effort indicators and the way they are evaluated can indeed help them to assess the learners and the reasons for their successes or failures. Second, a dashboard can make it possible for teachers to get feedback from their learners about the use of the resources available on the VLE. Teachers thus know which activities are motivating or not, the time of the day when the exercises are done, the progress of the learners during the previous weeks and months, whether the resources have been downloaded, etc. Last, dashboards can help teachers to differentiate their learners' activities, through personalised activity recommendations for learners. One concern is that the indicators in the dashboard may contradict the teachers' assessment, who may perceive this contradiction as a questioning of their ability to evaluate learners. Moreover, teachers were also concerned about the amount of educational tools available and the amount of time required to handle each one. Although they see the potential gain of time of such tools in theory, they are concerned that in practice a teacher dashboard would represent some additional effort.

During the prototype session, the teachers were provided with paper tablets and text/image labels (see Figure 4) and had to propose a dashboard layout prototype.

Several findings were identified. First, all the teachers were interested in including elements that would help them in their most laborious evaluation tasks. Especially, the French education system recently added the obligation for teachers to evaluate a specific set of skills. Hence, all of the participants included a simple line plot visualization of the evolution of these skills, both for the class and for each learner.
.

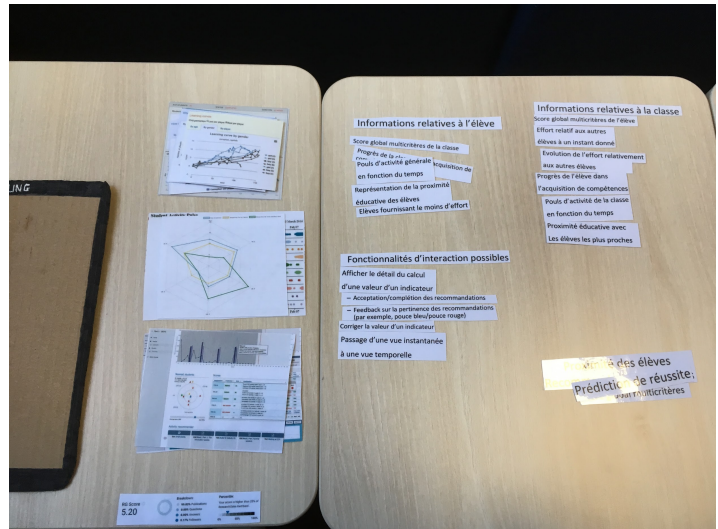

**Figure 4 Elements provided to the teachers for the layout prototyping session**

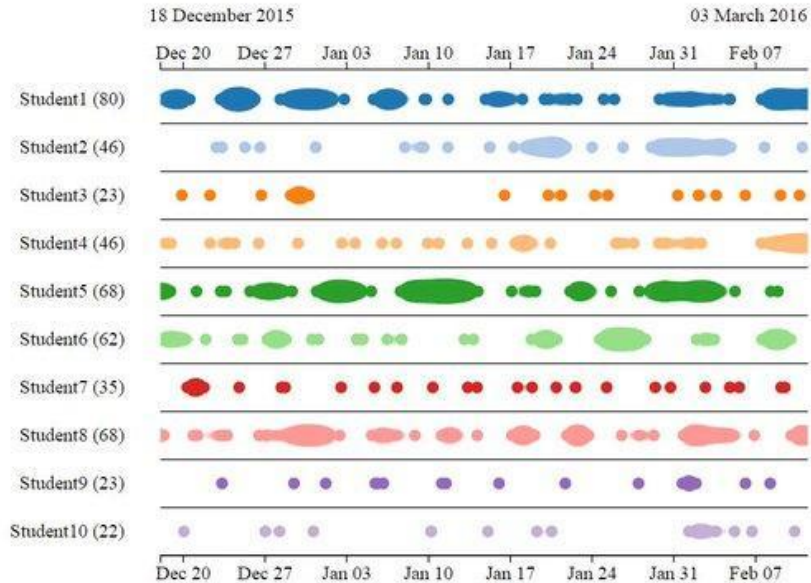

**Figure 5 Student activity pulse, from Workshops at LAK 2016**

Second, teachers put a strong emphasis on their desire for simplicity of the visualizations. Various "modern" visualizations were proposed, none of which were chosen by the teachers. This is in line with the desired low effort required to handle a new tool that was raised during the focus group. One exception is the image chosen by teachers to represent the learners' activity, which contains "pulses", i.e., ellipse representation whose size represents the learners' amount of activity (see Figure 5). It is not a traditional representation and seems to have seduced the teachers due to its ability to show detailed activity of learners in a unique component.

Last, all but one dashboard prototypes proposed by teachers include automated activity recommendations, together with the possibility for teachers to rate them. Most of the prototype dashboards also include the possibility for teachers to complete the proposed recommendations with additional activities to "help" the system to learn what to recommend. Based on these interesting findings, the final prototype dashboard has been shown to teachers, which do all agree on it. The development a first version is currently being conducted, the related results (in terms of acceptability, usability and utility) are therefore not available yet as experiments will be conducted in a few months, in parallel of experiments on learner dashboard.

*3.2 Dashboards for learners*

By providing learners with visual representations of their behavior or performance, dashboards are a way for these learners to be aware of their own performance or learning activity and to support their self-reflection. Dashboards can also be used to provide learners with information about others (behavior, performance, etc.), indicate whether their work is in-line with the expectation of the teacher, etc. Last, a dashboard can focus on the present, or on the evolution through time.

The choice of the information to display in a learner dashboard depends on several factors, as for the teacher dashboard.

The first factor is the available data. As previously said, the data correspond to learners' digital behaviour, or from information systems (see Section 2). The information, knowledge and indicators that we propose to mine from this data is the focus of section 5.

The second factor is related to learners' expectations. The information shown in the dashboard has to be useful to the learners (help them to support their self-reflection), and in some cases reaching their pedagogical goal (passing an examination).

The last factor is related to the display form that has to be comprehensible by the learners. This latter dimension is highly sensitive: whether the learners are K12 learners, under-graduate or graduate students, their level of understanding varies significantly as well as the way the information should be displayed. Several dashboards have been proposed in the literature, and mainly differ by their goal and the information they contain. The CALMsystem (Kerly, A., Ellis, R., & Bull, S., 2008) proposes a dashboard that displays the system estimated students level of knowledge on topics and students self-assessed confidence on these topics. In (Davis, S. K. et al., 2018) the dashboard displays undergraduate courses and aims at helping students to select some of them, etc. Most of the dashboards from the literature are dedicated to undergraduate or graduate students. As METAL is dedicated to K7 to K10 learners, these conclusions cannot fully considered.

In METAL, to comply with the profile of the targeted learners, a fourth factor is added: the understandability of the dashboard, relying on the hypothesis that most of the learners do not feel highly motivated by their learning, and that they do not want to spend time understanding the information contained in the dashboard. Thus, it has to be easy-to-access, so that all the learners quickly have an accurate idea of the information displayed. The main challenge here is thus not only to define understandable indicators but also the associated easy-to-understand visualization. The question: "are students able to interpret the information provided by such systems, and do they know what to do with it?" raised by (Corrin, L. & de Barba, P., 2015), is perfectly in line with our concerns. In addition, the conclusions drawn in several works such as (Ramos-Soto, A., Vázquez-Barreiros, B., Bugarín, A., Gewerc, A., & Barro, S., 2016), emphasize the fact that learners often misunderstand and misinterpret the information presented in the dashboards, which may have the exact opposite effect than expected. As their findings are about undergraduate students, they have to be considered cautiously in METAL. An important additional question in METAL is related to the impact of the dashboard on the concentration of students. (Lei, C.U. et al., 2018) revealed that some students can be distracted by the dashboard from their overall performance.

To design the learner dashboard, both teachers and learners play an important role. From the teacher dashboard co-design stage, some information has emerged about their vision of learners needs. They have mainly questioned the learners' acceptance of such a dashboard and pointed out the possibility that other stakeholders may also visualize the learner dashboard. This question is inline with (Whitmer, J., Nasiatka, D., & Harfield, T., 2018), that demonstrated a clear preference of learners for notifications that compare them to peers in their course, over notifications about trends in their own activity. This may encourage learners to accept that their own data is shared with other students. The question of sharing their data with teachers remains however open.

The co-design stage with learners is currently being set up. It will be conducted on 3 to 5 classes, with the goal to identify the information or indicator they expect to see or that they find useful for their self-reflection: learning progress, activity, information about their peers, the visualisation frequency of the dashboard (to avoid learners to be too much distracted), the possibility of sharing their data, and obviously the visual aspect of the indicators.

This co-design will be useful in several ways. First, it will provide a thorough understanding of the expectation of K7 to K10 learners, in terms of both indicators and visualisation. The question about the gap between their expectation those from undergraduate students will be answered. Second, the learners' feedback will help us to consider the relevance designing of several dashboards, with personalization seeks. This point of view has been studied in (Teasley, S.D., 2017) on undergraduate students, who showed that personalised designs have a positive effect on learners, including their motivation, especially when dashboards use social comparisons. Third, the co-design will help us consider the design of dashboards that propose recommendations of activities (as for the teacher dashboard): will learners be aware of such recommendations?

Once this co-design with classes will be conducted, the implementation of the learner dashboard can start. Meanwhile, few face-to-face interactions with learners have been conducted. First, but not significant conclusions can be drawn: learners are afraid about the comparison to their peers, that may demotivate them; learners do not expect temporal evolution of their activity or performance, they prefer a vision that represents the present moment.

## 4 Data Mining

With the increased availability of data, machine learning has become the main approach for knowledge acquisition and decision making systems, including in digital learning. It aims at designing a (mathematical) model of a phenomenon, relying on data. This data is supposed to represent, at least partially, this phenomenon.

Data Mining is a part of machine learning, and aims at extracting valuable and understandable information from large datasets, through the identification of regularities in data. The model of the phenomenon is made up of these regularities. Patterns are the most popular regularities studied in the literature. Their mining is one of the approaches used in METAL, due to their explainability characteristics, and are the focus of the first part of this section. A subsequent goal is to study the pertinence of using new types of data, such as learner gaze, to get information about learners' knowledge. It will be the focus of the second part.

*4.1 Multi-source Pattern Mining*

As introduced previously, data may come from several sources. It is the case in METAL, where data sources are manifold. Each source contains data about different dimensions of education and learning. For example, the *curriculum, demographic, teacher, school life, course, learner activity, educational resources*, etc. sources (these sources have been introduced in Section 2).

These sources contain data that reflect different learning dimensions and METAL assumes that mining all these sources together is a way to have a better understanding of learners learning activity. For example, the system can highlight positive correlations between the amount of activity of learners and their chances of success at exams, or identify mandatory resources.

In METAL, a pattern mining approach is adopted, more precisely a sequential pattern mining approach (Fournier-Viger, P., Lin, J. C. W., Kiran, R.U., Koh, Y.S., & Thomas, R., 2017), to discover information about learners. This approach has been chosen due to the understandability of the associated model. The challenge in METAL is the multi-source feature of data. Here is an example of a multi-source pattern we aim at discovering:

*{14-years, Male, Mathematics-grade-9} <R-15 R-42 R-Mathematics>*

This pattern is made up of two parts. The first one (between curly brackets) contains characteristics that learners must have: age (14-years) and gender (Male), mined from *demographic* source, and information from *curriculum* source (Mathematics-grade-9). The second part (between angle brackets) is a frequent sequential pattern that these learners often perform, mined from *learner activity* and *educational resource* sources: <R-15 R-42> is a sequence of two resource ids (from learner activity); followed by (R-Mathematics), an attribute of the resources commonly accessed afterwards, from *resource* source.

From our point of view, this pattern is richer than a pattern that would have been mined from the learner activity source only (that would only contain <R-15 R-42>). Indeed, it is at the same time more precise, in the sense that {14-years, Male, Mathematics-grade-9} acts as a precision about the type of learner that perform this pattern, and more "general" in the sense that the third element of the sequential pattern is more generic than a resource id. The pattern mining algorithm has been designed, it takes inspiration from PrefixSpan (Fournier-Viger, P. et al., 2017) and multi-relational pattern mining algorithms (Spyropoulou, E., De Bie, T., & Boley, M., 2014). To cope with the high complexity inherent to the mining of several sources, this algorithm proposes to give a priority to some sources, according to their nature and their links with other sources. As this algorithm is currently being evaluated, no insight of its efficiency can be drawn, except is ability to mine adaptable patterns. First patterns mined are quite intereseting and will be shown to teachers.

Mining such patterns has two main uses: patterns can be displayed in the learner or teacher dashboard, and rules can be inferred and recommendations of activities can be proposed to learners (or teachers) through the dashboard.

*4.2 Studying the link between gaze and memory*

As stated by (Klašnja-Milićević, Ivanović, Vesin, & Budimac, 2018), one of the main remaining challenges in e-learning is the integration of a factor that has been insufficiently considered in recommender systems: the student learning speed. Memory and learning are closely related notions because learners often have to memorize notions to learn concepts and acquire skills. For this reason, an additional focus of METAL is learners' memorization.

Teachers commonly evaluate what learners have memorized or learned through exercises or tests. Such an evaluation only allows periodic and partial validation of prior learning. Indeed, the proposed evaluations only cover an often small subset of concepts of the educational program. Moreover, the tests do not completely evaluate the memorization, because they mainly allow to evaluate the answers provided by the learner, not the approach that led to this result. It is therefore difficult for teachers to have an exhaustive view of learners' state of knowledge.

Based on these facts, the question is to know if it is possible to infer more precisely what learners remember and learn. To answer this question, METAL relies on learners' gaze data. Cameras and eye-trackers are now integrated in many devices and are promising tools to track learners' behavior and to model it as well as their understanding, in a more transparent and continuous way than periodic exams.

A learner gaze can be broken down into many features such as fixation points (coordinates of what the user has looked at on the screen), duration of the fixations, saccades (path between two fixations), angles formed by the scanpath or entropy level of the scanpath (Marchal, Castagnos, & Boyer, 2018).

The literature has reported that the gaze behavior could reflect what is on top of the cognitive processes. Especially,

(Glaholt, Wu, & Reingold, 2009) highlighted the fact that gaze can reveal the users' preferences, while interacting with IT tools. First attempts to establish a link between memory and gaze were initiated by (Ryan, J.D., Riggs, L, & McQuiggan, D.A, 2010). However, they found out that fixation points are insufficient to estimate memory state. At this point, let us note that there is actually not one single memory, but several forms of memory. As an example, (Clariana & Lee, 2001) distinguish recognition from recall. The recognition consists in remembering something when the stimuli is present, whereas the recall consists in remembering a stimulus which is not physically present. In METAL, it is being assumed that learning a concept consists in storing the information in the recall memory. While (Hannula, Baym, Warren, & Cohen, 2012) took an interest on the recognition memory, and (Bondareva et al., 2013) proved that gaze data efficiently predicts the quality of users' learning process, (Jackson, 2018) showed that eye gaze influences working memory in some conditions (e.g. visualizing happy faces, rather than angry faces). More recently, (Moissa, B., Bonnin, G., Castagnos, S., & Boyer, A., 2018) investigated the modelling of students' effort using gaze data. However, very few works focus on the link between recall and gaze.

In order to study and exploit this link between recall and gaze, the work first consists in verifying the existence of correlations between gaze features and recall memory, through user studies. A first study, conducted with 24 subjects who had to memorize 72 images, made it possible to establish statistically significant correlations between the recall memory, the number of fixations and the sum of the relative angles ($p < 2e-16$) (anonymous reference, 2016). A second study, conducted with 23 middle school students (11 females / 12 males) from 11 to 16 years old, confirmed the existence of correlations between memory and gaze features in the context of a language learning course. An ANOVA permutation test showed that what has been recalled during an exam after the learning phase is highly correlated with the scanpath length ($p = 0.0006$), the horizontal saccade amplitude ($p = 0.0002$), the sum of the relative angles in the scanpath ($p = 0.0034$) and the standard deviation of the relative angles ($p = 0.0004$).

The conclusion from these studies is that some gaze features are promising indicators of what has been recalled by learners. Since this work is still an ongoing research, it is now planned to design predictive models of the learners' state of memory and to integrate these inferences into the teacher and learner dashboards, so that learners and teachers have a more complete and regular view of their knowledge.

# 5   Conclusion

The focus of METAL is the design and development of Learning Analytics tools in secondary schools. It is at the same time an Open Learning Analytics initiative that aims at coping with the increasing complexity and great diversity in learning environments and a means to best support learners and teachers in improving the learning outcomes.

METAL aims at providing a concrete conceptual, developmental and operational plan to offer a framework to foster the dissemination of LA tools in secondary education. Challenges related to the conception, development and adoption are mainly due to the number of aspects to address.

First, METAL includes all the aspects of a LA environment: collection of multi-source data owned by a variety of stakeholders, selection and promotion of standards, design of an open-source LRS and the related legal and ethical issues, co-conception of dashboards with their final users to enhance acceptance, trust, usability, and design of explainable multi-source data-mining algorithms.

Second, METAL is also a research project with exploratory activities such as the estimation of the memorisation or the cognitive load via gaze; impact of LA tools on learners' motivation and engagement via an estimation of the cognitive load (ongoing work).

Last, METAL is also a project that assesses the impact of LA on learning outcomes via a longitudinal survey (ongoing work). METAL also provides solutions to the formation of teachers via a partnership with the training centre (initial training and Lifelong learning) for teachers, and actions of dissemination dedicated to the administrative staff and governance of secondary schools. All these aspects constitute the originality of METAL.

To date, the first results obtained, although incomplete, are promising. METAL shows that an open-source LRS can be designed and implemented, while promoting standards, even in the frame of multi-source data owners and data heterogeneity. METAL also shows that placing co-design at the heart of the development of dashboards, whether they are dedicated to learners or teachers, positively influences their design and first unexpected findings are already highlighted. Data mining algorithms can be designed in the frame of multi-source and heterogeneous data, and leads to a thorough view of the learning process.

Despite METAL is an ongoing project and findings are incomplete, it remains highly interesting as it embraces multiple aspects including technic, usability and impact. A strong point of METAL lies in the number of stakeholders that are involved (researchers, learners, teachers, institution, etc.) and that work together, which is of high complexity. METAL took up the challenge as shown by its smooth progress. It is an encouraging point for other future projects.

However, METAL faces some limits. First, the time required by design and development, due to the number of stakeholders, that are all involved in each step (it is a chosen strategy to ensure the final adoption of all the outputs). Second, the multidisciplinary aspect of the project has impact on its progress due to the numerous meetings and discussions conducted to ensure its success. Last, although all the stakeholders are involved each step of the project, many of them remain afraid by the project: teachers are afraid of having to modify their practices, parents are afraid that their children's' activities are collected, data owners are afraid of seeing their data stolen, etc. So, a significant amount of time is spent to reassure each of them.